\documentclass[aps,prb,twocolumn,amsmath,amssymb,nofootinbib,superscriptaddress,floatfix]{revtex4-1}

\usepackage{color}
\usepackage{xparse}
\usepackage{amsmath}
\usepackage{physics}
\usepackage{braket}
\usepackage{graphicx}
\usepackage{lipsum}
\usepackage{csquotes,helvet,mathpazo,listings}
\usepackage{notes2bib}
\usepackage{dcolumn}
\usepackage{bm}
\usepackage[normalem]{ulem}
\usepackage{verbatim}

\newcommand{\bs}[1]{{\boldsymbol{#1}}}

\begin{document}

\title{Quantum Fractality on the Surface of Topological Insulators}

\author{Lakshmi Pullasseri}
\affiliation
{
Department  of  Physics,  Emory  University,  400 Dowman Drive, Atlanta,  GA  30322,  USA
}

\author{Daniel Shaffer}
\affiliation
{
Department  of  Physics,  Emory  University,  400 Dowman Drive, Atlanta,  GA  30322,  USA
}

\author{Luiz H. Santos}
\affiliation
{
Department  of  Physics,  Emory  University,  400 Dowman Drive, Atlanta,  GA  30322,  USA
}

\date{\today}

\begin{abstract}
Three-dimensional topological insulators support gapless Dirac fermion surface states whose rich topological properties result from the interplay of symmetries and dimensionality. Their topological properties have been extensively studied in systems of integer spatial dimension but the prospect of these surface electrons arranging into structures of non-integer dimension like fractals remains unexplored. 
In this work, we investigate a new class of states arising from the coupling of surface Dirac fermions to a time-reversal symmetric fractal potential, which breaks translation symmetry while retaining self-similarity.
Employing large-scale exact diagonalization, scaling analysis of the inverse participation ratio, and the box-counting method, we establish the onset of self-similar Dirac fermions with fractal dimension for a symmetry-preserving surface potential with the geometry of a Sierpi\'nski carpet fractal with fractal dimension $D \approx 1.89$. 
Dirac fractal surface states open a fruitful avenue to explore exotic regimes of transport and quantum information storage in topological systems with fractal dimensionality.
\end{abstract}


\maketitle


\section{Introduction}

{Bloch's theorem, a cornerstone of our understanding of solid state behavior, dictates that single particle electronic states in regular crystals are organized into electronic bands. Bloch states also encode invaluable topological information that underlies the classification of free fermion symmetry-protected topological states. \cite{kitaev2009periodic,schnyder2009classification,ryu2010topological}
Deviations from perfect crystalline order are naturally present in any material, such as those produced by disorder and lattice defects. However, Bloch electrons remain a useful framework to characterize the low energy properties of weakly correlated materials when these effects are small.\cite{girvin2019modern}
On the other hand, it is fundamental to inquire what types of novel electronic states could arise when regular crystalline order is strongly violated, breaking away from the paradigm of Bloch states. Are there systematic ways to investigate electronic systems that lack translation invariance, but nevertheless retain some degree of structure beyond amorphous materials?

Fractals \cite{mandelbrot1982fractal,mandelbrot1967long} are a rich arena to explore this front. Interestingly, fractal geometries can lack translation symmetry while retaining self-similarity and point group symmetries, providing a promising framework to investigate emerging self-similar quantum states which depart from well-established properties of Bloch states. There is a long history of theoretical investigation of transport in fractal geometries, 
relating anomalous regimes of classical diffusion 
to the properties of the fractal,\cite{alexander1982density, blumen1984continuous, orbach1986dynamics, havlin1987diffusion} such as its fractal dimension. 
Furthermore,
experimental developments in recent years have brought renewed interest in the phenomenon of quantum fractality in electronic systems. 
For instance, the assembly of Sierpi\'nski gasket fractal networks on copper surfaces
via deposition of CO molecules produce surface states with fractal dimension.\cite{kempkes2019design}
Quantum states with fractal dimensionality open a venue to explore phenomena outside the realm of integer dimensions, allowing for novel regimes of transport and correlations. \cite{xu2021quantum,van2016quantum,bouzerar2020quantum,chamon2005quantum,nandkishore2019fractons,song2014topological,brzezinska2018topology,van2017optical,iliasov2020hall,westerhout2018plasmon,yang2020photonic,pai2019topological,biesenthal2022fractal} 
Moreover, they provide a rich setting to pursue new topological phenomena beyond topological Bloch states.

\begin{figure}
    \includegraphics[width = 8.5 cm]{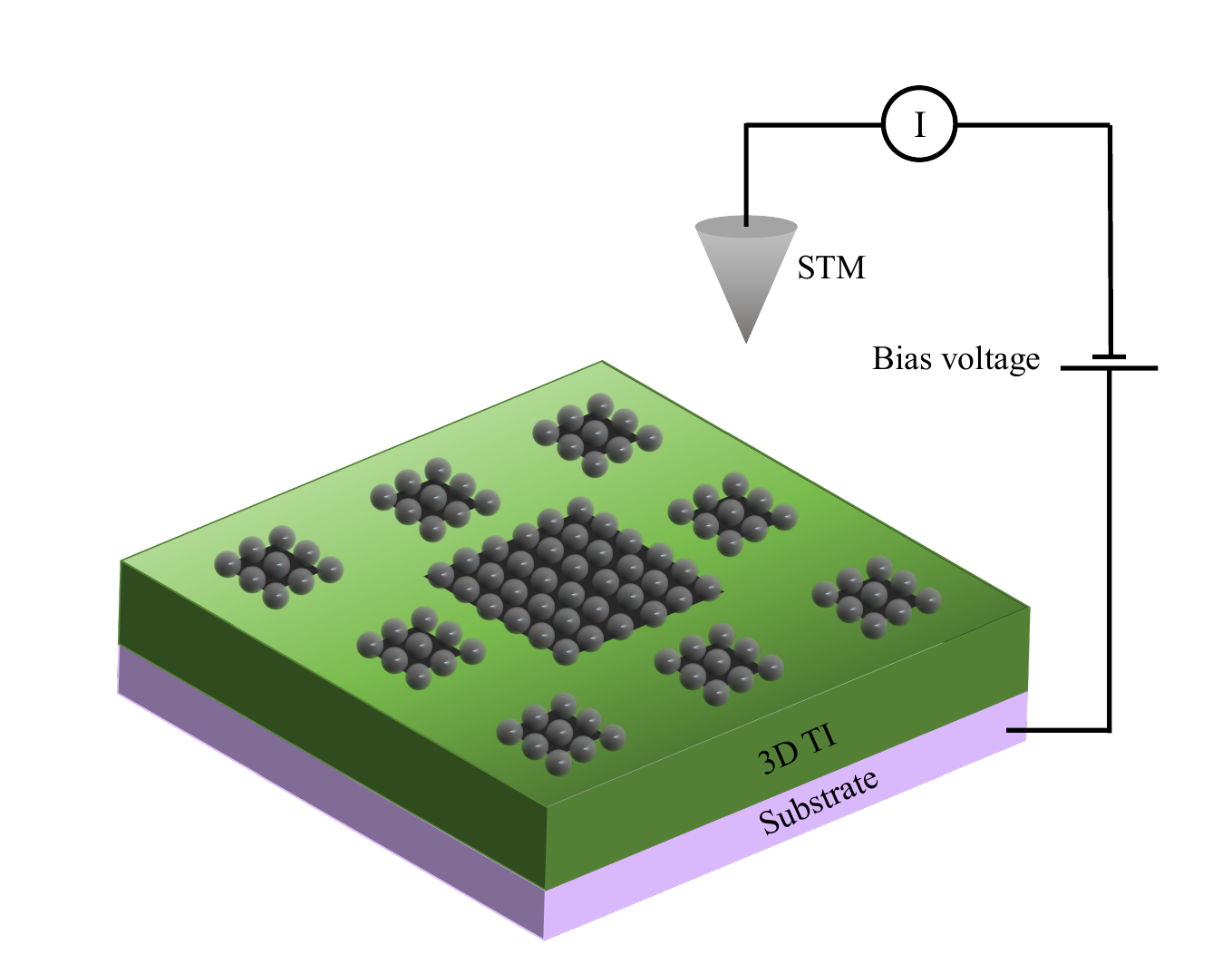}
    \caption{Schematic illustration of the experimental realization of the surface of a 3D TI coupled with a Sierpi\'nski carpet fractal. The experimental setup consists (from top to bottom) of a thin film of $3D$ TI grown on a substrate, scattering centers deposited on the $3D$ TI surface that lead to surface electronic states with
    Sierpi\'nski carpet fractal geometry, and STM apparatus to measure the LDOS of the surface states.}
    \label{fig: model}
\end{figure}

Motivated by these developments, in this {paper} we introduce and analyze a new fractal system formed on the surface of 3D time-reversal invariant topological insulators (TIs), \cite{moore2010birth,hasan2010colloquium,qi2011topological} in order to probe the response of surface Dirac fermions to a
scalar potential with fractal geometry, which respects the two key symmetries of 3D TIs, namely time-reversal and charge conservation symmetries. Unlike Schr\"odinger electrons, Dirac fermions on the surface of topological insulators are characterized by anomalous properties such as spin-momentum locking and nontrivial Berry phase. In this work, we analyze the electrons on the TI surface subject to a time-reversal invariant scalar potential $V(\bs{r})$ that couples to electron density according to the $H_{V} = \sum_{\sigma = \uparrow, \downarrow}\int d\bs{r} V(\bs{r})\,\psi^{\dagger}_{\sigma}(\bs{r}) \psi^{}_{\sigma}(\bs{r})$, and which has a self-similar fractal structure, as depicted in Fig.\ref{fig: model} for a Sierpi\'nski carpet. 

The fractal geometry could be imposed on the surface of $3D$ TI by atomic/molecular deposition techniques which have already been achieved on {copper surfaces}
creating unconventional states such as molecular graphene \cite{gomes2012designer} and electrons in a fractal Sierpi\'nski geometry. \cite{kempkes2019design} Nanopatterning \cite{forsythe2018band,dubey2013tunable,drienovsky2014towards} techniques that induce surface potential with fractal geometry could also open new pathways to design quantum states that combine topological and fractal properties. As we show in this work, emergent quantum fractality can be probed via scanning tunneling microscopy (STM) through distinct features in the local density of states (LDOS).

The lack of translation invariance combined with the presence of anomalous Dirac fermions on the TI surface makes their theoretical study nontrivial even at the non-interacting regime, which is the focus of this work. 
To address this new regime, we perform large-scale exact diagonalization (ED) using finite element discretization. 
While lattice discretization leads to fermion doubling, \cite{nielsen1981absence} we show in Sec.\ref{sec: model} that {the} mixing between the extra copies of Dirac fermions is negligible provided the fractal potential is smooth on the scale of the underlying lattice. This then provides a route to effectively model the response of a single Dirac cone on the TI surface to a fractal symmetry-preserving potential. Remarkably, our analysis uncovers Dirac quantum states that inherit the fractal dimension of the geometrical fractal deposited on the TI surface.

The focus of the present analysis
of Dirac electrons in fractal potentials without translation invariance 
goes beyond the recent works exploring the interplay of the Dirac fermions on the surface of $3D$ TI with a periodic lattice potential. \cite{wang2021moire,cano2021moire}
We also stress that this new regime is distinct from the well-explored multifractality studied in the context of localization transition in disordered electronic systems \cite{chamon1996localization,castillo1997exact,evers2008anderson,foster2012multifractal,foster2014topological,titum2015disorder,black2012strong,wu2014scattering,ziegler1996dirac,fedorenko2012two,wray2012topological,ludwig1994integer}  where the disorder induces multifractal scaling in the surface state wavefunctions at the localization threshold. Although disordered systems can support multifractal states that have a set of fractal dimensions, the possibility of Dirac surface states exhibiting fractal scaling behavior with a single scaling exponent that corresponds to a fractal geometry of choice has not been previously discussed, to the best of our knowledge. Establishing the existence of this new class of \textit{Dirac Fractals} on TI surfaces is the main result of this work. 
Additionally, our approach can be extended to 2D Dirac materials such as monolayer graphene and graphene-based heterostructures.

This paper is organized as follows. In Sec.~\ref{sec: model}, we introduce a model of the $3D$ TI surface coupled with {a} Sierp\'inski carpet fractal potential. In Sec.~\ref{sec: methods}, we discuss the results of the exact diagonalization of the model, and then numerically estimate the fractal dimension of the surface states by studying the scaling of the inverse participation ratio (IPR) with system size. The fractal dimension we obtain is further confirmed using the box-counting method.
We conclude with a discussion and outlook in Sec.~\ref{sec: conclusion}. 

\begin{figure*}
    \centering
    \includegraphics[width = 15 cm]{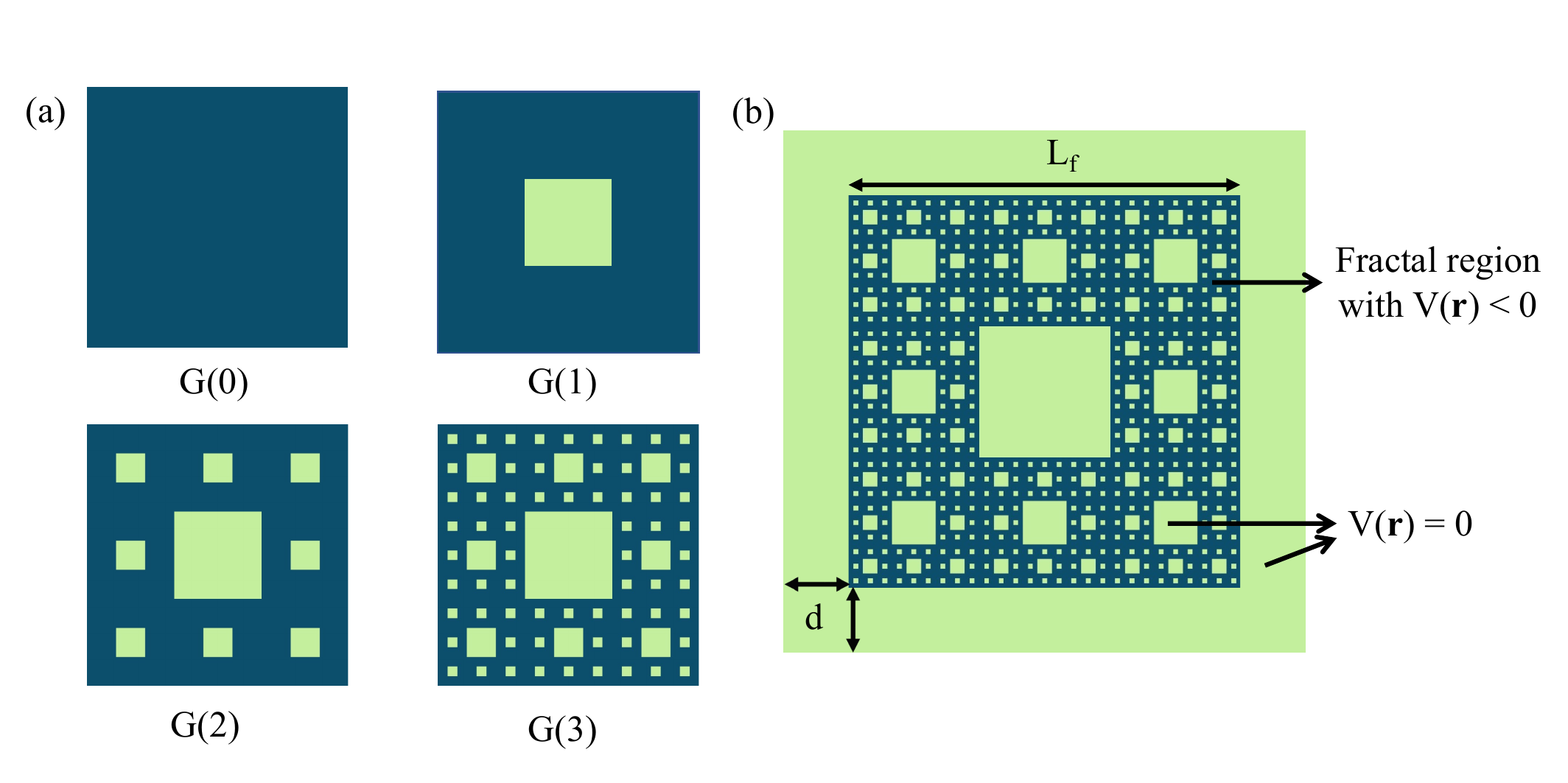}
    \caption{(a) Representation of the construction of the Sierpi\'nski carpet fractal. $G(0)$ is a $2D$ square that can be subdivided into 9 square units of the same size, from which the central unit is removed to obtain the $G(1)$ unit of the Sierpi\'nski carpet. 
    The same procedure is then applied to the remaining $8$ square units of $G(1)$ to generate $G(2)$. Similarly, $G(n)$ can be constructed from $G(0)$ by recursive application of this procedure.
    (b) Schematic of the model given in Eq. \ref{eq: model} that couples the surface Dirac cone of 3D TI with Sierpi\'nski carpet fractal of {fourth} generation. Here, the region on the 3D TI surface, shown in
    {a darker} color, which is the intersection of all the fractal units of generations ranging from {generations $1$ to $4$} is defined as the fractal region. It has a linear dimension $L_f$
    but has a Hausdorff dimension of 1.893.  
    The fractal potential $V(\bm{r})$ takes a non-zero value $V$ in the fractal region while zero otherwise. 
    In this model, the fractal region is positioned away from the boundary of the 3D TI surface by a finite distance, $\frac{d}{a} = 3$}.
    \label{fig: potential}
\end{figure*}

\begin{figure}
    \centering
    \includegraphics[width = 9.25 cm]{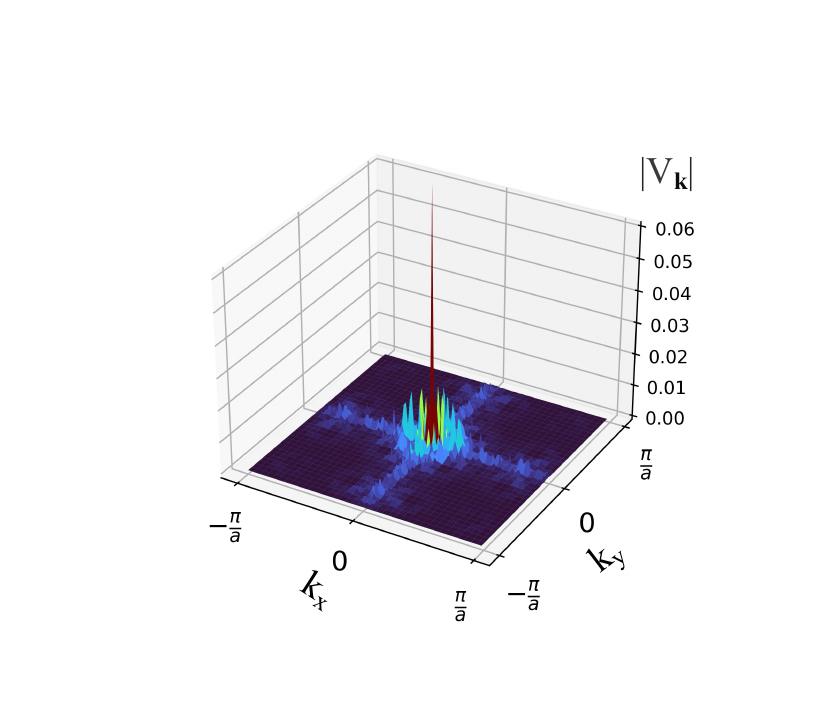}
    \caption{
    Fourier components ($V_{\bs{k}}$), in units of $t$, of the {$G(4)$} fractal potential given in Eq. \ref{eq: potential} with $V = -0.25\,t$, plotted in the first Brillouin zone of a $2D$ square lattice with lattice constant $a$. The dominating contribution comes from modes near zero momentum, with the largest red peak of height {$V_{\bs{k} = 0} = 0.14\,t$,} shown in red. {Note that the $z$-axis of the plot is cut off at $|V_{\bs{k}}| = 0.06\,t$ in order to display the details of the smaller peaks. The smallest length scale of $G(4)$ considered in this case is $l_4 = 3a$.} }
    \label{fig: VFourier}
\end{figure}

\section{Model}\label{sec: model}

In this section, we introduce a model of the 3D TI surface 
imprinted with a Sierpi\'nski carpet fractal geometry, as shown in Fig \ref{fig: model}. The model consists of the surface Dirac cone of the 3D TI, described by a continuum Dirac Hamiltonian, coupled with a time-reversal symmetric scalar potential $V(\bm{r})$. The resulting Hamiltonian takes the form
\begin{equation}
    \label{eq: model}
    H = \hbar v_F \big( \boldsymbol{\sigma} \times \bm{p} \big) + V(\bm{r}) \sigma_0
    \,,
\end{equation}
where $v_F$ is the Fermi velocity, $\bm{p} = \big(p_x,p_y \big)$ is the 2D momentum, $\boldsymbol{\sigma} = \big(\sigma_x,\sigma_y \big)$ are the Pauli matrices, and $\sigma_0$ is the identity matrix. Here, the scalar potential $V(\bm{r})$, also called the fractal potential, consists of a network of potential wells of strength $V$ arranged in a Sierpi\'nski carpet fractal geometry, as depicted in Fig.\ref{fig: potential}.

{The construction of the Sierpi\'nski carpet fractal starts with an \(L_f\times L_f\) square region that can be considered at the zeroth generation Sierpi\'nski carpet, denoted $G(0)$. The first generation carpet \(G(1)\) is obtained by dividing $G(0)$ into nine square sub-regions of size \(l_1=L_f/3\), and removing the central square, as shown in Fig.\ref{fig: potential} (a). The second generation carpet \(G(2)\) is then obtained by repeating the procedure on each of the remaining eight squares: each square is divided into nine squares of size \(l_2 = L_f/3^2\), and the central square is removed. In general, the \(G(n)\) generation carpet is obtained from the \(G(n-1)\) generation carpet by subdividing the \(l_{n-1}\)-sized squares from the previous step into nine squares of size \(l_n=L_f/3^n\), and removing all the central squares.
The Sierpi\'nski carpet fractal is obtained recursively as the limit of the sequence of \(G(n)\) as $n$ tends to infinity.
In practice we can only consider finite generation carpets as an approximation of the true fractal (up to \(n=4\) in our analysis), and take the potential $V(\bm{r})$ to have non-zero strength $V < 0$ only within the region \(G(n)\), as shown in Fig \ref{fig: potential} (b):
\begin{equation}
    \label{eq: potential}
    V(\bm{r}) = 
    \begin{cases}
        V < 0 & \bm{r} \in G(n)\\
        0 & \text{otherwise}
    \end{cases}
\end{equation}
and study the scaling properties of the system as \(n\) increases.}

{A defining characteristic of fractals is that they obey a scaling law characterized by a non-integer fractal dimension that remains constant over a range of length scales.\cite{mandelbrot1982fractal} 
In the simplest cases, the fractal dimension can be found using a box counting argument by considering the number \(N_n\) of boxes of linear size \(l_n\) needed to cover a region of total linear size \(L_f\). A two-dimensional \(L_f\times L_f\) square, for example, can be covered with \(N_n=(L_f/l_n)^2\) such boxes, while a three-dimensional cube can be covered with \((L_f/l_n)^3\) boxes. In general, \(N_n=(L_f/l_n)^D\) defines the Hausdorff dimension \(D\) of the set. Applying the box-counting method to the generation $n$ Sierpi\'nski carpet $G(n)$, since by construction \(G(n)\) consists of $N_n = 8^n$ squares of width 
\begin{equation}
    \label{eq: fractal_l}
    l_n = \frac{L_f}{3^n}
    \,,
\end{equation}
taking $N_n = \big(L_f/l_n \big)^D$ gives the Hausdorff dimension of the Sierpi\'nski carpet fractal as $D = \ln 8/ \ln 3 \approx 1.893$. We observe that for $G(n)$ with fixed finite $n$, the fractal scaling behavior will be seen for boxes of size \(l_n<l<l_1\), and no scaling behavior will be seen below the length scale \(l_n\).

Since the fractal potential lacks translational invariance, a representation in terms of Bloch states is impossible.
We, therefore, resort to an ED analysis of the single particle spectrum of the Hamiltonian Eq. \eqref{eq: model}, discretizing the Dirac model on a finite lattice that covers the fractal region. We chose a $249 \times 249$ square lattice with lattice constant $a=1$, on which it is simpler to define the Sierpi\'nski carpet fractal. The Dirac operator is discretized on this lattice using the finite difference method.
With this choice we are constrained to look only at the first four generations of the Sierpi\'nski carpet, lying within a square region of linear dimension $\frac{L_f}{a} = 243$ that we refer to as the fractal region below. Obtaining the full eigenspectrum of this model requires performing ED on an $N \times N$ lattice Hamiltonian matrix where $N = 2 \times 249^2 = 124002$ (the factor of $2$ accounting for spin). 
We work with boundary conditions that terminate the wavefunction at a distance $d \ll L_{f}$ away from the fractal region, as shown in Fig. \ref{fig: potential}.

In the absence of the fractal potential and with periodic boundary conditions, our choice of lattice regularization for the Dirac fermions leads to the periodic Bloch Hamiltonian
\begin{equation}
    \label{eq: modelPBC}
    H_{0} = t \big[ \sin (k_y a) \sigma_x - \sin (k_x a) \sigma_y \big]
    \,,
\end{equation} 
where $t = \frac{\hbar v_F}{a}$. This Hamiltonian supports four Dirac cones at momenta $\{ (0,0), (0,\pi/a), (\pi/a, 0), (\pi/a, \pi/a) \}$ in the Brillouin zone, each of which is described by Eq. \eqref{eq: model} with $V = 0$.
Since the regime of linearly dispersing Dirac fermions occurs in the energy domain $|E| \lesssim t$, we restrict the analysis of the effect of the fractal potential onto states within an effective bandwidth $W \sim 2t$. Furthermore, we require that $W/V \lesssim 1$, so that the fractal potential does not exceed the energy scale of the 3D TI bulk energy gap.

The appearance of an even number of Dirac cones in the lattice Hamiltonian
$H_0$ is a familiar manifestation of fermion doubling.\cite{nielsen1981absence}
In order to capture the properties of the TI surface described by a single component Dirac fermion, we restrict the fractal potential to have negligible Fourier components with $|\bs{k}| \gg \frac{1}{a}$, which ensures that the coupling between different Dirac cones is negligible. 
\begin{figure*}
    \centering
    \includegraphics[width = 17.5 cm]{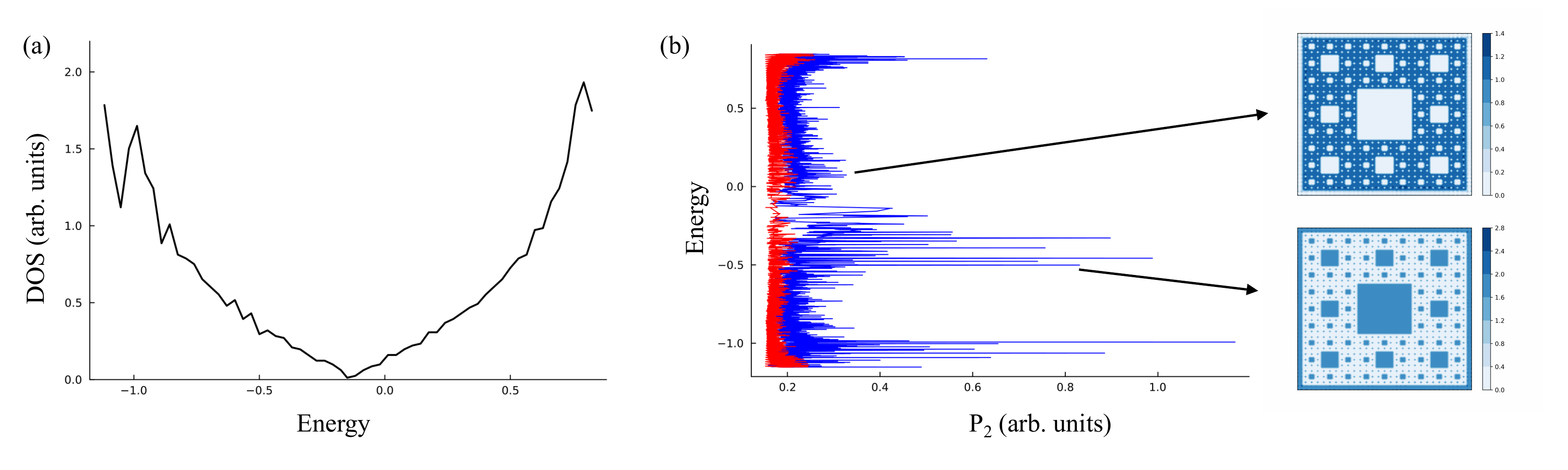}
    \caption{(a) DOS plotted vs energy, measured in units of $t = \frac{\hbar v_f}{a}$, for a $G(4)$ fractal potential of strength $V = -0.25\,t$. The DOS vanishes at $E_D = -0.14\,t$, and the considered energy window is centered around $E = E_D$ with a width of $W = 2t$. (b) The $q=2$ moment of IPR ($P_2$) of each individual state, as given by Eq. \eqref{eq: ipr}, plotted against energy,  measured in units of $t = \frac{\hbar v_f}{a}$, for {$G(4)$ fractal potential of strengths $V=-0.25\,t$ (blue) and $V=0$ (red)}. The LDOS maps of the states {corresponding to $V=-0.25\,t$,} that lie within an energy window of width $\Delta E = 0.01\,t$, around $E = 0.1\,t > E_D$ and $E = -0.53\,t < E_D$, as indicated by the arrows, are shown on the right.}
    \label{fig: dos}
\end{figure*}
We verified this condition for fractal potentials smooth on the scale of the discrete lattice i.e, $l_n \gg a$, as depicted in Fig. \ref{fig: VFourier} that shows that the dominant Fourier components of the fractal potential are peaked at small momenta $|\bs{k}| \ll \frac{1}{a}$.
In addition, the inversion symmetry of the Sierpi\'nski carpet forces the Fourier components with
momenta $\bs{k} \in \{ (0,\pi/a), (\pi/a, 0), (\pi/a, \pi/a) \}$ to vanish, thus further suppressing the coupling between different Dirac cones.
\section{Methods and Results}\label{sec: methods}

In this section,  we discuss the results of the ED analysis 
of the model given in Eq. \eqref{eq: model} and investigate the emergence of the Dirac surface states with fractal dimensionality in the presence of the potential $V(\bm{r})$. 
We identify the dimensions of the states using two complementary approaches.
In Sec.~\ref{subsec: IPR}
the fractal scaling behavior of the surface states is numerically established by analyzing
the distribution of the IPR of the single-particle states, and in Sec.~\ref{subsec: BoxCounting} we confirm these results using the standard box-counting applied to the local density of states. In both approaches, we numerically observe the occurrence of states with fractal dimension $\approx 1.89$.

As we are regularizing the single Dirac cone using a square lattice with lattice constant $a$ and hopping parameter $t$, 
the interplay between the fractal potential and kinetic effects is captured by the dimensionless parameter $V/t$. 
The ED analysis \cite{sakurai2003projection} of the model is performed with a fractal potential of strength $V/t = -0.25$, which corresponds to $12.5 \%$ of the effective bandwidth $W$, for each of the first four generations of the Sierpi\'nski carpet, and the obtained eigenspectra are studied to seek out the onset of fractality in the Dirac surface states. 

The large-scale diagonalization yields information about the single particle eigenstates formed in the fractal region over a wide range of energies. With this dense data, we perform several lines of inquiry, starting with the analysis of the density of states (DOS). We observe that the DOS corresponding to a $G(4)$ fractal potential shown in Fig \ref{fig: dos} (a), vanishes at non-zero energy $E_D \approx -0.14 t$. This implies that the fractal potential shifts the massless Dirac cone in energy to $E_D = V_{\mathbf{k}=0}$
while leaving the time-reversal symmetry (TRS) intact, so that the spectrum is formed by degenerate Kramers pairs. 
We numerically confirm that the shift in energy $E_D$ is directly proportional to $V$ such that the Dirac point moves down (up) in energy as $V$ grows more negative (positive). {Additionally, we note that the particle hole symmetry of the DOS about \(E_D\) is broken when the potential is introduced.}

To probe the localization properties of eigenstates $\psi(\bm{r})$, we employ the inverse participation ratio, \cite{wegner1980inverse} $IPR_q = \sum_{i} |\psi(\bm{r}_i)|^{2 q}$ where $i$ is summed over all the lattice points of the system and {$q \in \mathbb{R}$}. A uniform state has an $IPR_2 = 1/N$ where \(N\) is the total number of sites, {and in general \(1/IPR_2\) can be considered as an effective number of sites \(N_f\) to which the wave function is localized.
For a state with an effective fractal dimension, we expect the same scaling for \(N_f\) as for the number of boxes needed to cover the fractal, i.e. \(N_f\sim(L_f/l)^D\) where \(l\) is the lattice constant. However, as discussed in Sec. \ref{sec: model}, for a finite generation approximation of the fractal \(G(n)\) the fractal scaling persists only for boxes of size \(l_n\) and above. To take this into account, we consider a coarse-grained IPR obtained by}  
averaging the wave function over the length scale $l_n$, dividing the $L_f$ sized square region shown in Fig. \ref{fig: potential}(a) into $l_{n}$-sized regions $\mathcal{A}_{i}$, each of which is treated as a coarse-grained lattice site. 
The coarse-grained probability distribution is then defined as $|\bar{\psi}_{i}|^2 = \sum_{j \in \mathcal{A}_{i}}|\psi(\bm{r}_j)|^{2}$ and the corresponding coarse-grained IPR is given by
\begin{equation}
    \label{eq: ipr}
    P_q = \sum_{i} |\bar{\psi}_{i}|^{2q} 
\end{equation}
where the index $i$ runs over all the the $l_{n}$-sized regions denoted by $\mathcal{A}_{i}$. Wave function normalization ensures $P_{q=1} = 1,$ and henceforth we focus on {$q > 1$} modes.

In Fig. \ref{fig: dos}(b), we display the second moment of the IPR {($P_2$)} numerically calculated for each of the eigenstates corresponding to the $G(4)$ fractal potential. The spectrum consists of Kramers pairs of localized states characterized by high IPR peaks in an energy window of width $W = 2t$ around $E_D$.
{A striking feature is the appearance of two classes of states separated by the scale $E_D$.}
{The manifold of states with $E > E_D$ exhibits a characteristic LDOS distribution consistent with the fractal, as shown in the top right LDOS map of Fig. \ref{fig: dos} (b), {where the probability of finding electrons outside the fractal region, though non-zero, is very small.}  
{Importantly, not all the individual eigenstates belonging to this class of states above $E_D$ exhibit this characteristic LDOS distribution with very small support from outside the fractal region. As the wide range of heights shown by their corresponding IPR peaks in Fig. \ref{fig: dos} (b) indicates, the individual eigenstates have different IPR values, thereby different spatial distributions of the LDOS. But interestingly, when we consider the LDOS map averaged over in energy that corresponds to the states in an energy window of width $\Delta E \sim 10^{-2}\,t$, we find the emergence of the characteristic LDOS profile shown in the top right LDOS map of Fig. \ref{fig: dos} (b), that mirrors features of the Sierpi\'nski carpet fractal. This suggests that most of the individual eigenstates above $E_D$, in an energy window of width $\Delta E$ feature the geometry of the Sierpi\'nski carpet, as we confirm below.}
%

The states with $E < E_D$, on the other hand, show more pronounced IPR peaks, with spatial profile shown by the bottom right LDOS map of Fig. \ref{fig: dos} (b). These correspond to electronic configurations with probability dominated by contributions outside the fractal region. Such high IPR peaks can be attributed to the fact that the area outside the fractal region is smaller than the area of the fractal region when approximated by finite generation sets \(G(n)\) for $n\leq 4$; for sufficiently large \(n\) we, therefore, expect the IPR of states with \(E>E_D\) to eventually exceed that of states with \(E<E_D\). Already at \(n\leq4\), we observe additional lower IPR peaks for \(E<E_D\), that correspond to more delocalized states that have a substantial weight in the fractal region in comparison to the states with higher IPR peaks.
In all cases, the states with \(E<E_D\) lack self-similarity, indicating the absence of fractal scaling behavior. Importantly, we have observed the same pattern described above for all four fractal generations of the Sierpi\'nski carpet fractal potential that we studied.}

While the LDOS maps provide important visual guidance of the onset of Dirac states with fractal dimension, we confirm that the states with energy $E > E_D$ indeed inherit the scaling properties of the Sierpi\'nski carpet fractal by calculating their fractal dimensions through IPR scaling in Sec.\ref{subsec: IPR}, and box-counting methods in Sec.\ref{subsec: BoxCounting}.

\subsection{Scaling of the IPR logarithm}\label{subsec: IPR}

\begin{figure*}
    \centering
    \includegraphics[width = 18 cm]{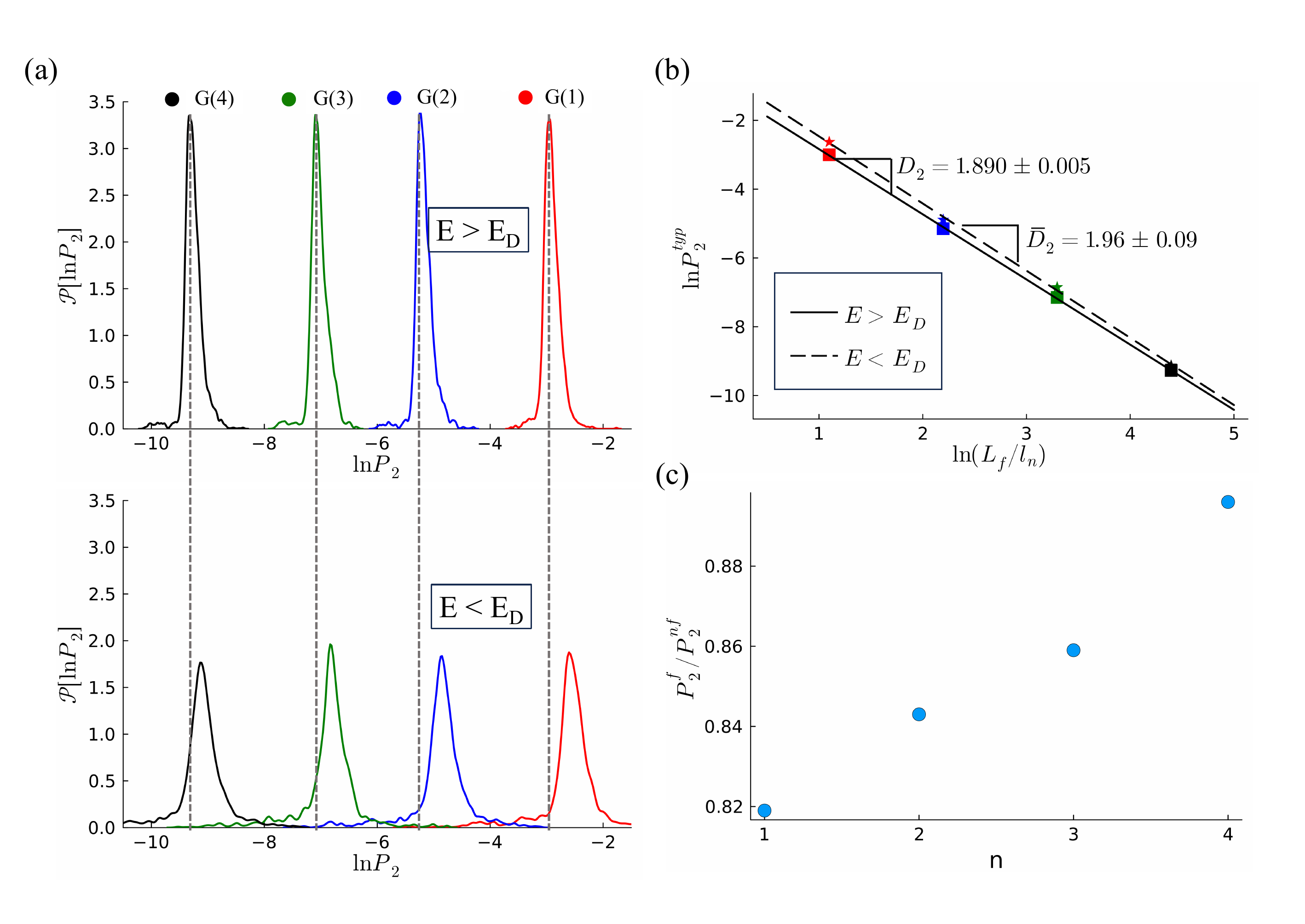}
    \caption{(a) Distribution of the IPR logarithm, $\mathcal{P}[\ln P_2]$ of the two classes of states separated by $E_D$, for each of the first four generations of the Sierpi\'nski carpet fractal potential of strength $V = -0.25\,t$. The dashed line indicates the mean $\ln P^{typ}_2$ of $\mathcal{P}[\ln P_2]$ corresponding to $E > E_D$ for each of the first four fractal generations. 
    (b) The $\log$-$\log$ scaling of the typical values $\ln P^{typ}_2$ shown in (a), with linear dimension $L_f/l_n = 3^n$. The fractal dimensions, $D_2$ and $\bar{D}_2$ of the set of states with $E > E_D$ and $E < E_D$ respectively, are given by the corresponding regression slopes.
    (c) Ratio of the typical IPR values ($P^{typ}$) corresponding to $E > E_D$ and $E < E_D$, denoted as $P^{f}_2$ and $P^{nf}_2$ respectively, plotted as a function of the fractal generation $n$.}
    \label{fig: ipr}
\end{figure*}

{In this section, we provide numerical evidence that states with energy $E > E_D$ have a fractal dimension, using the scaling of the coarse-grained IPR distribution. As alluded to above, the inverse of the second moment IPR \(P_2\sim 1/N_f\) can be considered as the effective number of coarse-grained sites occupied by the eigenstate. If the eigenstate is completely uniform within the fractal region and vanishes outside of it, the IPR coarse-grained over boxes of size \(l_n\) would then be given by precisely \(1/N_n\) with \(N_n\) being equal to the number of boxes of size \(l_n\) needed to cover the fractal region. In that case, we, therefore, conclude that \(P_2 = (L_f/l_n)^{-D}\) where \(D\) is the fractal dimension of the Sierpi\'nski carpet obtained using the box-counting method, and which can therefore alternatively be obtained by scaling \(P_2\) with the generation \(n\).

In general, the wave functions are non-uniform and non-vanishing, but since \(P_q\) remains well-defined we can obtain an effective dimension of the wave functions by considering how \(P_q\) scales with the generation of the fractal. If the wave functions in a given energy window are scale-invariant on a range of length scales \(l\) with \(a\ll l\ll L_f\), the IPR \emph{distribution} \(\mathcal{P}\) over a set of states has a universal form independent of system size and box size \(l_n\).\cite{wegner1980inverse,huckestein1995scaling,evers2000fluctuations} More precisely, the distribution \(\mathcal{P}\left[\ln (P_q/P_q^{typ})\right]\) of the IPR logarithms scaled by a typical IPR \(P_q^{typ}\) (which can be taken to be the mean of the IPR distribution) is independent of \(L_f\) and \(l_n\). The scaling properties of the IPR are contained in \(P_q^{typ}\), which in this satisfies
\begin{equation}
    P^{typ}_q \propto \left( \frac{L_f}{l_n} \right)^{-D_q(q-1)}
    \label{eq: IPRscaling}
    \,,
\end{equation}
where $D_q$ is the fractal dimension that can be different for different $q$, which signals multifractality.\cite{mandelbrot1974intermittent, hentschel1983infinite} If \(D_q=D\) for all \(q\), the system is fractal and there is no multifractality.}

To test if this scaling scenario holds for states with $E > E_D$, as suggested by the characteristic LDOS maps shown in Fig. \ref{fig: dos}(b), we plot the $\ln P_2$ distribution for the states with $E > E_D$ for the first four generations of the fractal potential, as displayed at the top of Fig. \ref{fig: ipr}(a). We note that each of the distributions displays a sharp peak and narrow width, indicating that most of the states above $E_D$ have similar {spatial support.}
Moreover, the distributions appear similar for all the fractal generations, and we confirm that they largely coincide upon shifting along the horizontal axis. The same behavior is also exhibited by $\mathcal{P}[\ln P_3]$.

In Fig. \ref{fig: ipr}(b) we show the scaling of the corresponding typical values $P^{typ}_2(n)$ (given by the mean value of the distribution) with the linear dimension $L_f/l_n$, and extract the fractal dimension $D_q$ from the scaling exponent according to Eq.\eqref{eq: IPRscaling}. \cite{fyodorov1995mesoscopic,mirlin2000multifractality,evers2000fluctuations,prigodin1998long} From the regression slopes, we determine that the states with $E > E_D$ have fractal dimension $D_2 = 1.890 \pm 0.005$, which is very close to the dimension of the Sierpi\'nski carpet.  
{We also check for possible multifractal behavior\cite{mandelbrot1974intermittent} by calculating the $q=3$ as well as $q=1.5$ scaling exponents, which gives the corresponding fractal dimensions, $D_3 = 1.86 \pm 0.04$ and $D_{1.5} = 1.895 \pm 0.008$, as shown in Appendix \ref{App: IPR}.}
We are unable to check the scaling behavior for {$q > 3 $} due to system size limitations. However, since the fractal dimension of the Sierpi\'nski carpet falls within the confidence interval of the estimated value of {$D_2$, $D_{1.5}$ and $D_3$}, our findings strongly suggest that for $E > E_D$, the states on the surface of the 3D time-reversal invariant topological insulator are well localized in the fractal, thus inheriting the fractal dimension of the Sierpi\'nski carpet.

{We similarly carry out the same IPR scaling analysis for states with $E < E_D$. The corresponding $\ln P_2$ distributions for the first four generations of the fractal potential are plotted in Fig. \ref{fig: ipr}(a). In this case, we do not expect a scaling behavior for low generations if the wave functions are localized to the non-fractal region, because the box-counting dimension of the non-fractal region converges to two only for sufficiently large generations \(n\gg 4\), as can be verified by direct computation. Surprisingly, we nevertheless find that the scaling of the typical IPR values, again given by the mean of the distribution, with $L_f/l_n$ results in a scaling dimension $\bar{D}_2 = 1.96 \pm 0.09$, as shown in Fig. \ref{fig: ipr}(b). This suggests that the typical wave functions for states with $E < E_D$ are close to two dimensional already for small generations, despite the presence of states with high IPR peaks located on the tails of the distributions shown at the bottom of Fig. \ref{fig: ipr} (a) which appear to be localized to the non-fractal region.
This suggests that these typical states have wave functions that leak considerably into the fractal region. This additionally implies that although the IPR of the fractal states with \(E>E_D\) is on average lower than that of the non-fractal states with \(E<E_D\), as the generation increases the IPR of the fractal states will eventually exceed that of the non-fractal states. This trend can be seen by comparing peaks of the IPR distribution for fractal and non-fractal states, as shown by vertical dashed lines in Fig. \ref{fig: ipr}(a): the non-fractal distribution can be seen to shift to the left (towards lower IPR) faster than the fractal distributions as the generation increases. This trend is explicitly verified in Fig. \ref{fig: ipr}(c).

We also observe that the IPR distributions for states with $E < E_D$ are broader than for states with $E > E_D$, which reflects the wider range of IPR peaks for $E < E_D$ shown in Fig. \ref{fig: dos}(b). 
The states localized to the non-fractal region corresponding to the highest IPR peaks, which we do not expect to exhibit two-dimensional scaling at low generations, lie in the higher tail ends of the IPR distributions, so these are not typical states. For high generations \(n\gg4\), we expect the dimension of these atypical states to tend to two and their IPR to decrease as the area outside to the area inside the fractal grows with $n$.}

\subsection{Box-counting method}\label{subsec: BoxCounting}

\begin{figure}
    \centering
    \includegraphics[width = 8.5 cm]{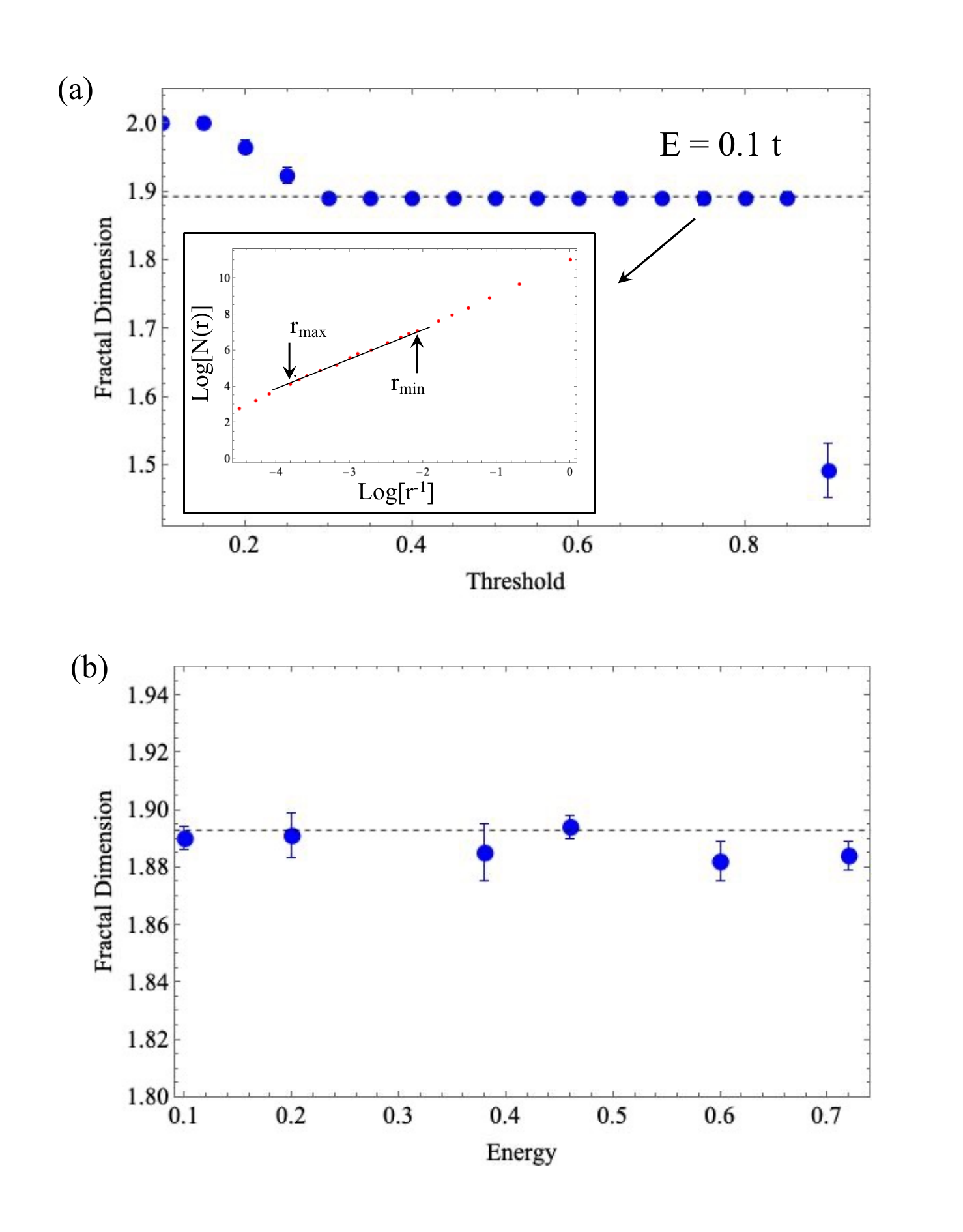}
    \caption{(a) Fractal dimension $D$ of the states within an energy window of width $\Delta E = 0.01 t$ around $E = 0.1\,t$, for a $G(4)$ fractal potential of strength $V = -0.25\,t$, calculated using the box-counting method, plotted against the binarization threshold, $\alpha$. The plateau at $D=1.890 \pm 0.004$ indicates the fractal dimension for threshold values ranging from $0.3$ to $0.85$. Note that the error bars of the plot are smaller in size compared to the scale of the y-axis.
    Inset shows the log-log plot of $N(r)$ vs $1/r$ corresponding to the threshold $\alpha = 0.75$. Here, $r_{min}$ and $r_{max}$, respectively, denote the minimum and maximum box sizes chosen to estimate the box-counting dimension of the states.
    (b) Fractal dimensions of the states as a function of the corresponding energy values for a $G(4)$ Sierpi\'nski carpet fractal potential of strength $V = -0.25\,t$, calculated using the box-counting method, as shown in (a). The black dashed line in (a) and (b) marks the fractal dimension of the Sierpi\'nski carpet at $D = 1.893$.}
    \label{fig: BoxCounting}
\end{figure}

In this section, we provide an alternative confirmation of the fractal dimension of states with energy $E > E_D$ using the box-counting method applied to the LDOS. The method is based on counting the number $N(r)$ of boxes  of size $r$ that cover the LDOS profile of the wavefunction, leading to the Minkowski-Bouligand (or the box-counting) dimension\cite{mandelbrot1985self}
\begin{equation} 
\label{eq: box counting D}
    D = \lim_{r \rightarrow 0} \frac{\ln (N(r))}{\ln (1/r)}
    \,.
\end{equation}

To binarize the LDOS map we introduce a threshold $\alpha > 0$, such that values of LDOS below (above) $\alpha$ are set to zero (one) in the binarized map. Varying \(\alpha\) allows us to probe whether the fractal dimension obtained via Eq.~\eqref{eq: box counting D} is stable over a range of threshold values. A threshold that is too low or too high makes the binarized map sensitive to very short distance features of the LDOS, for instance, due to the lattice discretization. Additionally, the scaling given by Eq.~\eqref{eq: box counting D} holds\cite{bouda2016box,foroutan1999advances} only in a finite range of box sizes $r \in [r_{min}, r_{max}]$, as it breaks down when box sizes are either too close to the linear dimension of the system (this tends to overestimate the dimension), or too small and pick up spurious short distance behavior due to noise. Following Ref.~\onlinecite{foroutan1999advances}, we determine this finite range as follows. We start with all the possible box sizes $r$. As we reduce $r$ from the maximum possible box size, we obtain a regression slope of two until $r$ reaches $r_{max}$. Then $r_{min}$ is determined as the smallest box size that gives a linear regression (with a maximum standard deviation of $10^{-3}$) in the range $r \in [r_{min}, r_{max}]$.

In Fig. \ref{fig: BoxCounting}(a), we show the Minkowski-Bouligand dimension for an energy window of width $\Delta E = 0.01\,t$ around $E = 0.1\,t > E_D$ for a $G(4)$ fractal potential $V = -0.25\,t$, and for different values of the binarization threshold ranging from $0$ to $1$, where the LDOS values are rescaled such that their minimum and maximum values are $0$ and $1$ respectively. 
When the threshold $\alpha$ is set close to zero such that we pick up contributions from even the shorter LDOS peaks, the LDOS landscape appears homogeneous, thereby resulting in a box-counting dimension of $two$. On increasing $\alpha$ further, interestingly, we find a range of threshold $\alpha \in [0.3, 0.85]$ that shows the emergence of a set of dominant LDOS peaks spread across the fractal region, as indicated by the plateau at $D = 1.890 \pm 0.004 $, which characterizes the onset of fractal dimensionality in good approximation with that of the Sierpi\'nski carpet fractal.
Importantly, our approach reveals the stability of the fractal scaling behavior over a considerable range of threshold values.

Moreover, the box-counting method applied to a larger set of states with $E > E_D$ reveals, as shown in Fig. \ref{fig: BoxCounting}(b), that their fractal dimension is in good agreement with the fractal dimension of the Sierpi\'nski carpet. 
Therefore, combined with the analysis of the IPR scaling discussed in Sec. \ref{subsec: IPR}, the fractal dimension of the TI surface state obtained for the Minkowski-Bouligand dimension provide strong evidence that the Dirac surface states can inherit the scaling properties of the Sierpi\'nski carpet fractal. Furthermore, we have observed similar fractal dimensions for another value of the fractal potential with $V=-0.5 t$, suggesting that the observed fractal character of the states is robust against varying the strength of the fractal potential.

\section{Discussion and Outlook} \label{sec: conclusion}

Time reversal invariant 3D topological insulators are a large
class of symmetry-protected topological states with surface Dirac fermions protected by time-reversal and charge conservation symmetries, which
provide a rich arena to explore exotic regimes of quantum matter.  
In this work, we {used large-scale exact diagonalization to study} a new class of Dirac surface states formed when a symmetry preserving one-body potential with the geometry of a fractal is superimposed on the surface of this topological material, opening a new direction to interrogate and explore topological fermionic states with fractal properties.

The main finding of this study is the onset of Dirac fractals, gapless  Dirac states whose wave function acquires the fractal dimension of the applied potential over a wide range of energies. These results were obtained for a symmetry-preserving one-body fractal potential having the geometry of a Sierpi\'nski carpet fractal. We have employed two methods based on inverse participation ratio as well as {box counting}, which {show within numerical accuracy} that Dirac fermions on the TI surface acquire the fractal dimension of Sierpi\'nski carpet fractal $D \approx 1.89$. Furthermore, our analysis does not support the scenario of multifractality. Therefore, {to our knowledge,} this work presents the first numerical evidence of time-reversal invariant Dirac surface states {of} fractal character, which expands the realm of fractal quantum states beyond time-reversal broken fractal Hofstadter states. \cite{Hofstadter76,wang_classification_2020,Shaffer2022Unconventional}

This research opens an interesting avenue to search for exotic fractal quantum orders on the surface of topological insulators. A fruitful direction would be to characterize new regimes of charge and energy transport in such fractal networks for the Sierpi\'nski carpet and beyond. Moreover, probing the response of such states perturbations that break time-reversal and/or charge conservation symmetry symmetries
could offer new ways to understand fractionalization phenomena in the fractal setting. 
{Another direction would be to understand the formation of fractal states in Weyl semimetals and, in this context, to understand the role of weak disorder on surface states.
\cite{slager2017dissolution,wilson2018surface}}
The advent of new experimental 
platforms that could induce a surface potential with fractal geometry using techniques such as nanopatterning \cite{forsythe2018band,dubey2013tunable,drienovsky2014towards}, and molecular deposition \cite{gomes2012designer,kempkes2019design}, {combined} with probe techniques such as scanning tunneling microscopy, offer
promising experimental landscapes to realize and probe Dirac fractals.

\section*{Acknowledgments}
We thank Ribhu Kaul, Gil Refael, Rafael Ribeiro, 
Ajit Srivastava and Daniel Sussman for insightful discussions. This research was supported by the U.S. Department of Energy, Office of Science, Basic Energy
Sciences, under Award DE-SC0023327. Numerical analysis in this work was partially carried out on the Indiana Jetstream2 cluster with support from the NSF ACCESS program. {L. P. acknowledges funding from the Women in Natural Science Fellowship of Emory University.}

\appendix

\section{Confirming the absence of multifractality}\label{App: IPR}

\begin{figure}
    \centering
    \includegraphics[width = 8.5 cm]{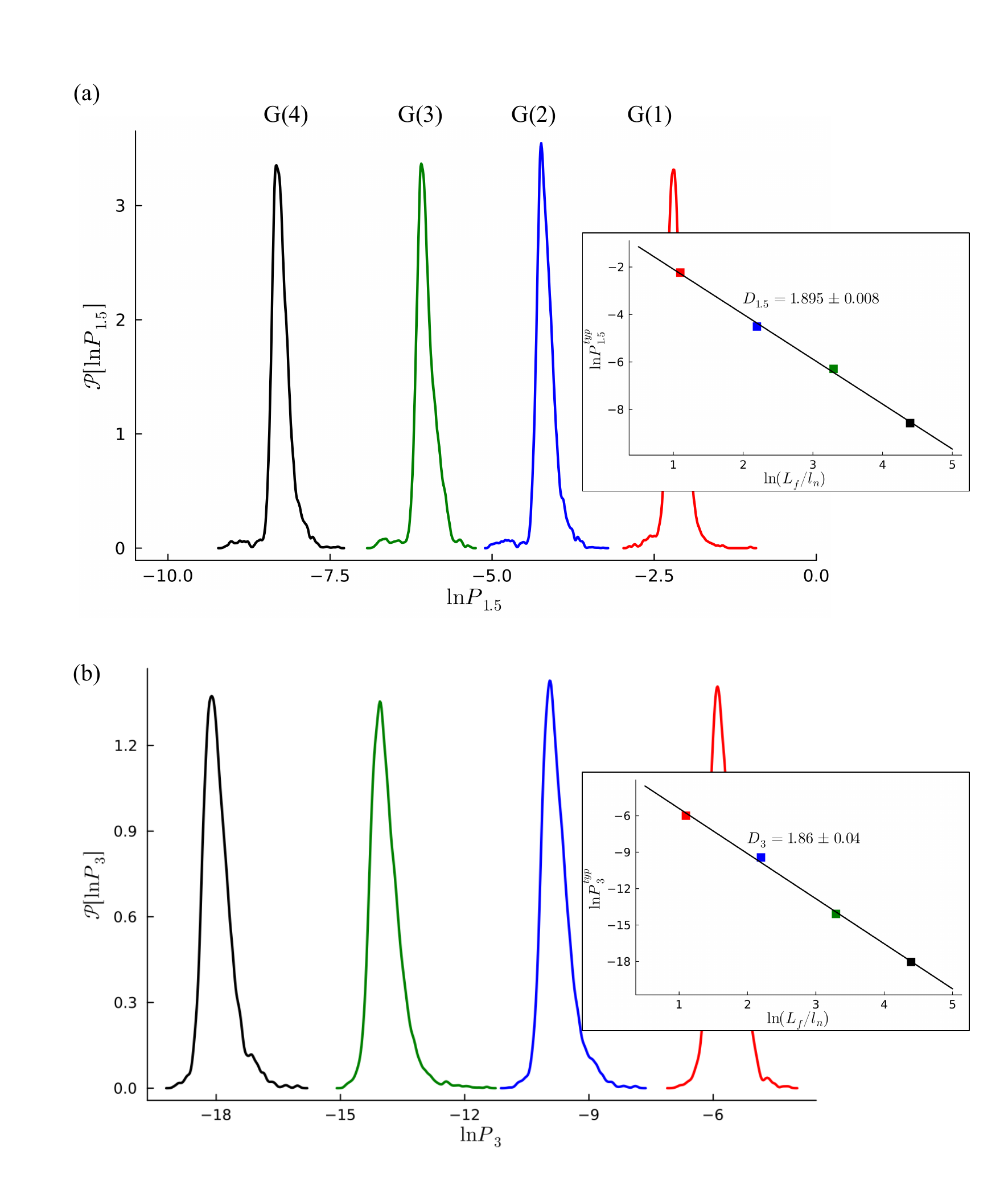}
    \caption{Distribution of the IPR logarithm, $\mathcal{P}[\ln P_{q}]$ of the class of states above $E_D$, for each of the first four generations of the Sierpi\'nski carpet fractal potential of strength $V = -0.25\,t$, for (a) $q = 1.5$, (b) $q = 3$. The $\log$-$\log$ scaling of the corresponding typical values $\ln P^{typ}_{q}$ with the linear dimension $L_f/l_n = 3^n$ are shown in the inset, where the scaling dimensions $D_{1.5}$ and $D_3$ are given by the corresponding regression slopes.}
    \label{fig: App_ipr}
\end{figure}

As shown in Fig. \ref{fig: ipr}, the $q=2$ scaling dimension $D_q$ of the manifold of Dirac surface states above the shifted Dirac point $E_D$ is very close to the fractal dimension of the Sierpi\'nski carpet. While it strongly suggests that the surface states exhibit fractal scaling behavior for $q=2$, it does not rule out the possibility of multifractality where the scaling exponent $D_q$ varies with $q$. Hence, we calculate the scaling exponents for $q=1.5$ and $q=3$, and we obtain $D_3 = 1.86 \pm 0.04$ and $D_{1.5} = 1.895 \pm 0.008$, as shown in Fig. \ref{fig: App_ipr}. We can notice that $D_2 \approx D_{1.5} \approx D_3$, which strongly indicates the absence of multifractal behavior.

\bibliographystyle{apsrev4-1} 
\bibliography{apssamp}

\end{document}